\documentstyle[aps,preprint,epsf]{revtex}

\newcommand{\be}[1]{\begin{equation}\label{#1}}
\newcommand{\ee}{\end{equation}}

\newcommand{\prlput}[1]{}
\newcommand{\rem}[1]{}
\newcommand{\FFig}[1]{Fig.~{\ref{fig:#1}}}
\newcommand{\eref}[1]{(\ref{#1})}
\newcommand{\Eref}[1]{Eq.~(\ref{#1})}

\newcommand{\NN}{{\cal N}}
\def\rarr{\rightarrow}
\def\eps{\epsilon}

\def\tanb{\tan\beta}
\newcommand\tanbt[1]{\tan^{#1}\beta}
\def\tc{\theta_c}
\def\cd{{c\dagger}}

\begin{document}
\draft
\pagestyle{empty}
{\tightenlines

\preprint{
\begin{minipage}[t]{3in}
\begin{flushright}
IASSNS--HEP--98/74\\OSU-HEP-98-5\\
hep-ph/9808447 \\
\end{flushright}
\end{minipage}
}

\title{\Large\bf A Reexamination of Proton Decay \\
in Supersymmetric Grand Unified Theories}
\author{{\bf
K.S. Babu$^{1,2}$} and
{\bf
Matthew J. Strassler$^1$}}
\address{\ \\ $~^1$School of Natural Sciences,
Institute for Advanced Study,
Olden Lane,
Princeton, NJ 08540, USA\\}
\address{\ \\ $~^2$Department of Physics, Oklahoma State
University, Stillwater, OK 74078, USA.\footnote{Permanent address
as of August 1, 1998.}
\\ ~ \\
{\tt babu@osuunx.ucc.okstate.edu, strasslr@ias.edu}}
\maketitle
\begin{abstract}

We reconsider dimension--five proton decay operators, making
semi-quantitative remarks which apply to a large class of
supersymmetric GUTs in which the short-distance operators are
correlated with the fermion Yukawa couplings. In these GUTs, which
include minimal $SU(5)$, the operators
$(u^cd_i^c)^\dagger(d_j\nu_\tau)$, induced by charged Higgsino
dressing, completely dominate for moderate to large $\tanb$.  The rate
for $p \rarr (K^+,\pi^+) \overline{\nu}_\tau$ grows rapidly, as
$\tanbt4$, and the $K^+$ to $\pi^+$ branching ratio can often be
precisely predicted.  At small $\tanb$ the operators $(u d_i)(d_j
\nu)$ are dominant, while the operators $(u d_i)(u \ell^-)$, with
left-handed charged leptons, are comparable to the neutrino operators
in the generic GUT and suppressed in minimal GUTs.  Charged-lepton
branching fractions are always small at large $\tanb$.  The electron
to muon ratio is small in minimal GUTs but can be larger, even of
order one, in other models.  All other operators are very small.  At
small $\tanb$ in non-minimal GUTs, gluino and neutralino dressing
effects on neutrino rates are not negligible.
\end{abstract}
}


\newpage
\pagestyle{plain}

\vskip 0.4 in

\draft
\tightenlines

Unification of the three gauge forces of the standard model into a
single symmetry group is an attractive idea with a long history
\cite{patisalamA,patisalamB,ggGUT,gqwGUT,sakaisGUT,dgsGUT}.  Grand
Unified Theories (GUTs) provide a beautiful explanation of the
multiplet structure of the standard model fermions, and predict the
value of one of the gauge couplings at the weak scale in terms of the
other two.  The success of this prediction in the context of the
minimal supersymmetric extension of the standard model (MSSM) has
given new impetus to the study of SUSY GUTs over the past decade; for
a review see \cite{susyunifrev}.

 GUTs place quarks, leptons and their antiparticles in the same
multiplet, thereby making the nucleon unstable.  The momentum scale of
grand unification, as inferred from the extrapolation of the standard
model gauge couplings, is $M_{\rm GUT} \sim 2 \times 10^{16}$ GeV, so
large that nucleon decay rates are compatible with present
experimental limits.  In supersymmetric GUTs there are two sources of
baryon number violation. Dimension--six (four--fermion) operators are
generated by the exchange of superheavy gauge bosons of the GUT
symmetry group.  With $M_{\rm GUT} \sim 2 \times 10^{16}$ GeV, the
expected proton lifetime from these operators is $\tau(p \rightarrow
\pi^0 e^+) \simeq 10^{36 \pm 2}$ yr.~\cite{susyunifrev,hmy,nathC}; the
present experimental lower limit is $\tau(p \rightarrow \pi^0 e^+)\ge
2 \times 10^{33}$ yr.~\cite{superpi}.  The second source of proton
decay --- the subject of this paper --- is the dimension--five
(two--fermion/two--sfermion) operators induced by the exchange of
color--triplet Higgsinos, the GUT partners of the MSSM Higgs(ino)
doublets \cite{sakai,weinbergsusy}.  The associated amplitudes scale
as $M_{\rm GUT}^{-1}$, but are suppressed by light fermion Yukawa
couplings.  They are also suppressed by a loop factor, which arises
from the ``dressing'' of the operator by a gaugino or Higgsino that
converts the two sfermions into light fermions.  The size of the loop
integral is highly uncertain, due to the unknown masses of the
supersymmetric particles.  Additional uncertainty stems from the
unknown values of $\tanb \equiv \left\langle H_u
\right\rangle/\left\langle H_d\right \rangle$ and the triplet Higgsino
mass.  (In branching ratios, some of the unknown parameters cancel, so
they can be more reliably predicted than the overall rate.)  Since the
strength of these operators is proportional to Yukawa couplings, the
dominant modes are those with strange quarks -- the heaviest fermion
into which the proton can decay kinematically.  In minimal SUSY
$SU(5)$, including all the uncertainties, the lifetime may be
estimated as $\tau(p \rightarrow K^+ \overline{\nu}) =
(10^{28}-10^{34})$ yr.  This range is nearly eliminated by the
experimental limit $\tau(p \rightarrow K^+ \overline{\nu}) \ge 5
\times 10^{32}$ yr. \cite{superk}, suggesting that proton decay must
be discovered imminently if the idea of grand unification is on the
right track.

These issues have been studied extensively in minimal $SU(5)$
\cite{sakai,weinbergsusy,hmy,nathC,raby,rudaz,vysotskii,pal,nath,nathB,goto}
--- see \cite{hmy,nathC} for more references to the early literature
--- as well as in other GUTs \cite{bbtwo,bbsotenA,nathD,lucrab,macph}.
Most calculations have been specific to particular models.  In this
letter we attempt to make statements which, while more qualitative
than quantitative, apply to a wide class of theories.  Our purpose in
this paper is to summarize and clarify the expectations concerning
branching ratios in a large ensemble of supersymmetric GUTs, in which
the dimension-five operators are roughly correlated with the fermion
Yukawa couplings.  We will make this more definite below.  Our
approach is somewhat similar in spirit to that of \cite{kaplan}.

General statements of this type require a thorough analysis.
To this end a systematic method 
has been developed in which rough upper bounds on
baryon-number-violating operators may be established; this will be
presented elsewhere \cite{estimates}.  These bounds apply under the
following conditions.  The dimension-five operators take the form
\be{Lfiveo}
{\cal L}_5 = \int d^2\theta
\left\{\lambda_{ijkr}Q_i Q_j Q_k L_r
+\bar\lambda_{ijkr} U^c_iD^c_r U^c_j E^c_k\right\}+h.c.
\ee
(Capital letters denote superfields; $U_i$ contains the squark $\tilde
u_i$ and the quark $u_i$, {\it etc.})  We take as our ansatz the
relations $\lambda_{ijkr}\sim\bar\lambda_{ijkr}\sim{1\over
M}\eps_i\eps_j\eps_k$, independent of the index $r$, where $M$ is an
overall mass scale independent of $i,j,k,r$ and $\eps_3\sim 1,
\eps_2\equiv \eps \sim 1/25$ and $\eps_1\sim\eps_2^2$.  (The motivation
\cite{babubarr,msfermion} for the ansatz is associated with the
fact that, near the GUT scale, most of the flavor structure of the
theory can be roughly encoded in powers of $\eps$ \cite{tenc}.)
In \eref{Lfiveo} the $\lambda_{ijkr}$ are expressed in the gauge
basis, but it will be more convenient to work in the ``supersymmetric
basis'', in which the matter fermions are expressed as mass
eigenstates and the sfermions are expressed as the
supersymmetric partners of those eigenstates.  To convert from the
gauge $(g)$ to the supersymmetric $(m)$ basis, we must rotate the
superfields as $U_i^{(g)} =
R^u_{ij}U_j^{(m)},~D_i^{(g)}=R^d_{ij}D_j^{(m)},$ {\it etc.} so
that the fermion mass matrices are diagonal. 
In the supersymmetric basis,
\begin{equation}\label{Wfivesusy}
W_{dim-5}=(\hat{\lambda}^\ell_{ijkr} U_i^\alpha D_j^\beta
U_k^\gamma E_r
-\hat{\lambda}^\nu_{ijkr}
U_i^\alpha D_j^\beta D_k^\gamma
\NN_r
+\hat{\bar{\lambda}}_{ijkr} \bar{U}_i^\alpha \bar{D}_r^\beta
\bar{U}_j^\gamma \bar{E}_k) \epsilon_{\alpha\beta\gamma}~.
\end{equation}
where $\alpha,\beta,\gamma$ are color indices,  we have dropped the
superscript $(m)$, and we have defined
\begin{eqnarray}\label{lambdahat}
\hat{\lambda}^\ell_{ijkr} &=&\sum_{i',j',k',r'=1}^3
\lambda_{i'j'k'r'}R^u_{ii'}R^d_{jj'}R^u_{kk'}R^e_{rr'} \ ; 
\end{eqnarray}
the coefficients $\hat\lambda^\nu_{ijkr}$ and
$\hat{\bar{\lambda}}_{ijkr}$ are defined analogously.
As part of our ansatz, we assume the matrices $R^u$, $R^d$,
$R^{\bar u}$, $R^{\bar e}$ --- those associated with the ${\bf 10}$
representations of $SU(5)$ --- have the same texture as
$V_{CKM}$; we do not assume this for the other $R$ matrices.  If
$V_{CKM}$ satisfied $V_{ij} \sim \min\{\eps_i/\eps_j,\eps_j/\eps_i\}$, then
we would have $\hat{\lambda}^\ell_{ijkr} \sim \hat{\lambda}^\nu_{ijkr}
\sim \hat{\bar{\lambda}}_{ijkr} \sim {\lambda}_{ijkr} \sim
\eps_i\eps_j\eps_k $.  However, the Cabibbo angle $\tc$ is of order
$5\eps_1/\eps_2$ and can enter into the $\hat\lambda$. An enhancement
by a factor of $\theta_c/({\eps_1\over\eps_2})\sim 5$ may occur for
each index $i,j,k$ taking value 1 \cite{tenc}; thus
\be{ansatz}
\hat{\lambda}^\ell_{ijkr} \sim \hat{\lambda}^\nu_{ijkr}
\sim \hat{\bar{\lambda}}_{ijkr}
\sim \eps_i\eps_j\eps_k
\Big[\theta_c/ (\eps_1/\eps_2)
\Big]^{(\delta_{i1}+\delta_{j1}+\delta_{k1})} \ .
\ee
Note $\delta_{i1}+\delta_{j1}+\delta_{k1}\leq 2$ because of the
antisymmetry properties of the $\hat\lambda$ coefficients.  More
details of this ansatz are given in \cite{estimates} and \cite{tenc}.

If the coefficients $\hat\lambda$ in a given GUT are all of order or
less than those appearing in \Eref{ansatz}, then we can apply the
results of \cite{estimates} to the model.  Although we will not
classify them here, many GUTs (including minimal $SU(5)$ and $SO(10)$
and their more realistic variants, as well as the ten-centered models
of \cite{babubarr,msfermion,tenc}) are compatible with the ansatz
\eref{ansatz}.  We will now demonstrate that the $\hat\lambda$
coefficients in the minimal $SU(5)$ GUTs are consistent with
\eref{ansatz}.  In the supersymmetric basis, the superpotential induced
by the color--triplet Higgsinos in minimal $SU(5)$ is
\be{GUTops}
W_C =  (U_i D_j e^{i \sigma_i} f_i V_{ij} + U_i^c E_j^c f_i V_{ij})H_C
+ (D_i \NN_i h_i - U_i E_j V_{ij}^* h_j + U_i^c D_j^c e^{-i
\sigma_i}V_{ij}^* h_j)\bar H_C~.
\ee
Here $H_C$ and $\bar H_C$ are the color triplet and anti--triplet from
the ${\bf 5}_H$ and ${\bf \overline{5}}_H$ of the Higgs multiplet, $V$
is the CKM matrix, $f_i$ ($h_i$) is the diagonal Yukawa coupling
matrix of the up--quarks (down-quarks and leptons), and the $\sigma_i$
are phase factors with $\sum_{i=1}^3 \sigma_i = 0$.  Exchange of the
Higgs color triplets, of mass $M$, leads to the dimension-five operators
\begin{eqnarray}\label{dimfive}
W_{d=5} &=& M^{-1}(U_i D_j D_k \NN_r e^{i \sigma_i} f_i V_{ij}
\delta_{kr} h_r - U_i D_j U_k E_r e^{i \sigma_i} f_i V_{ij} V_{kr}^*
h_r + U_i^c D_r^c U_j^c E_k^c e^{-i\sigma_i} f_j V_{jk} V_{ir}^* h_r)
\end{eqnarray}
along with other terms irrelevant for proton decay.

If it were the case that $\tc\sim\eps_1/\eps_2$, so
that $V_{ij}\sim \min\{\eps_i/\eps_j,\eps_j/\eps_i\}$, along with 
$f_i\sim \eps_i^2$, $h_i\sim \eps_i\zeta$, where $\zeta=\tanb/60$, 
we would have
\begin{eqnarray}\label{GUTmatches}
  f_i V_{ij} \delta_{kr} h_r \sim \eps_i^2 \eps_r \delta_{kr}
\min\{\eps_i/\eps_j,\eps_j/\eps_i\}\zeta \alt \eps_i\eps_j\eps_k\zeta
\nonumber \\
f_i V_{ij} V_{kr} h_r \sim \eps_i^2 \eps_r
\min\{\eps_i/\eps_j,\eps_j/\eps_i\}\min\{\eps_k/\eps_r,\eps_r/\eps_k\}\zeta
\alt \eps_i\eps_j\eps_k\zeta
\nonumber \\
 f_j V_{jk} V_{ir}^* h_r \sim \eps_j^2 \eps_r
\min\{\eps_j/\eps_k,\eps_k/\eps_j\}\min\{\eps_i/\eps_r,\eps_r/\eps_i\}\zeta
\alt \eps_i\eps_j\eps_k\zeta
\end{eqnarray}
All coefficients in \eref{GUTmatches} would then be equal to or
smaller than those in \Eref{ansatz}; note the overall factor of
$\zeta$ can be absorbed into the overall mass scale which affects all
$\hat\lambda$ equally.  Now let us account for the fact that $\tc\sim
5\eps_1/\eps_2$.  Inspection of \Eref{GUTmatches} shows that factors
of the Cabibbo angle can only enhance a coupling above
$\eps_i\eps_j\eps_k$ if one of the indices $i,j,k$ takes value 1;
$r=1$ cannot cause such an enhancement because of the factor $h_1$,
which reduces $\hat\lambda_{ijk1}^{(e,\nu)},\hat{\bar\lambda_{ijk1}}$
below its ansatz value.  If two such indices take value 1, then one
may get a double enhancement.  The resulting coefficients are the same
as or less than those appearing in \Eref{ansatz}, and thus the upper
bounds of \cite{estimates} are applicable.

Rough upper bounds from \cite{estimates} on various four-fermion
baryon-number-violating operators are given in the Table.  All
operators not shown are negligibly small. The Table shows the
different contributions from various gauginos and Higgsinos.
Wino-Higgsino mixing is not listed; in each case the upper bound on
such contributions lies between or below the pure Wino and the pure
Higgsino bounds.  Explicitly indicated are factors of $\eps$,
$\theta_c$, gauge coupling constants $g_i$ and third-generation Yukawa
couplings $y_b,y_\tau$ which stem from left-right mixing or Higgsino
couplings; note $y_b$ and $y_\tau$ are roughly of order $\zeta$.  Also
appearing are parameters $\gamma$ and $\delta$.  The first measures
the extent to which the $\tilde b,\tilde t$ are split in mass from and
mixed with the other left-handed squarks; if the messenger scale of
supersymmetry breaking is high, then $\gamma\sim 1$, while if it is
near the weak scale $\gamma$ may be small.  The factor $\delta =
(A\cos\beta+\mu\sin\beta)(A\sin\beta+\mu\cos\beta)v/\tilde m^2$
parameterizes the size of left-right squark mixing; here $v,\mu,\tilde
m,A$ are $246$ GeV, the Higgsino mass, the universal squark mass and
the coefficient of the trilinear scalar terms.  An overall factor of
$M^{-1}\zeta$ is omitted from every entry.  Neutrino flavors have not
been distinguished, and the flavor of a charged lepton is only
indicated when the bounds depend on the lepton flavor.  For additional
details see \cite{estimates}.

\bigskip

\def\udsn{$uds\nu$}
\def\uusl{$uus\ell$}
\def\uddn{$udd\nu$}
\def\uudl{$uud\ell$}
\def\ucdcsn{$(u^c d^c)^\dagger s\nu$}
\def\ucscdn{$(u^c s^c)^\dagger d\nu$}
\def\ucdcdn{$(u^c d^c)^\dagger d\nu$}
\def\ucmucsu{$ (u^c \mu^c)^\dagger su$}
\def\ucmucdu{$(u^c \mu^c)^\dagger du$}
\def\ucecsu{$(u^c e^c)^\dagger su$}
\def\ucecdu{$ (u^c e^c)^\dagger du$}

\begin{tabular}{|l|c|c|c|c|c|}
\hline
&&$\tilde W^{\pm}$&$\tilde H^{\pm}$&$\tilde g$&$\tilde W^0,\tilde B$
\\ \hline
\udsn,\uddn&&$g_2^2(\eps^3\tc^2,\eps^3\tc^3)
$&$\delta y_b^2(\eps^4\tc,\eps^5\tc)
$&$g_3^2\gamma(\eps^3\tc^2,\eps^4\tc^2)
$&$g_2^2(\eps^3\tc^2,\gamma\eps^4\tc^2)$
\\ \hline
\uusl,\uudl&&$g_2^2(\eps^3\tc^2,\eps^3\tc^3)
$&$\delta y_b^2(\eps^4\tc,\eps^5\tc)
$&$g_3^2\gamma y_b^2(\eps^4\tc,\eps^4\tc^2)
$&$g_2^2\gamma y_b^2(\eps^4\tc,\eps^4\tc^2)$
\\ \hline
\ucdcsn&&$g_2^2\delta y_\tau\eps^2\tc$&$y_\tau\eps^2\tc$&$0$&$0$
\\ \hline
\ucscdn,\ucdcdn&&$g_2^2\delta y_\tau\eps^3\tc$&$y_\tau\eps^3\tc$&$0$&$0$
\\ \hline
\ucmucsu,\ucmucdu&&$g_2^2\delta y_b(\eps^4\tc^2,\eps^5\tc^2)
$&$y_b(\eps^4\tc^2,\eps^5\tc^2)
$&$g_3^2\gamma \delta y_b^3(\eps^5\tc,\eps^6\tc)
$&$g_2^2\gamma \delta y_b^3(\eps^5\tc,\eps^6\tc)$\\ \hline
\ucecsu,\ucecdu&&$g_2^2\delta y_b(\eps^4\tc^3,\eps^5\tc^3)
$&$y_b(\eps^4\tc^3,\eps^5\tc^3)
$&$g_3^2\gamma \delta y_b^3(\eps^5\tc^2,\eps^6\tc^2)
$&$g_2^2\gamma \delta y_b^3(\eps^5\tc^2,\eps^6\tc^2)$
\\ \hline
\end{tabular}

\vskip .2 in  {\it Rule-of-thumb upper bounds on coefficients
of four-fermion baryon-number-violating operators, as found in
\cite{estimates} using the ansatz in \Eref{ansatz}.   Important loop
factors, matrix elements and overall coefficients are omitted and must
be accounted for when using this Table.  See the text for further
explanation.}
\vskip .2 in

At small $\tanb$ the operators $uds\nu$, $uus\mu$ and $uuse$ have the
largest bounds, while at large $\tanb$ the operator
$(u^cd^c)^\dagger(s\nu)$ potentially exceeds all others.  The
replacement of the $s$ quark by a $d$ quark engenders a small
suppression \cite{raby,rudaz}.  As we will see, it is always true that
many of these operators saturate or nearly saturate their bounds.  It
follows from the Table that operators with right-handed leptons, even
$u^\cd \mu^\cd s u$, are negligible at all values of $\tanb$, as are
operators with a left-handed charged lepton and two right-handed
quarks.  We will therefore restrict out attention to the first four
rows in the Table.

\begin{figure}
\centering
\epsfxsize=3.2in
\hspace*{0in}\vspace*{.2in}
\epsffile{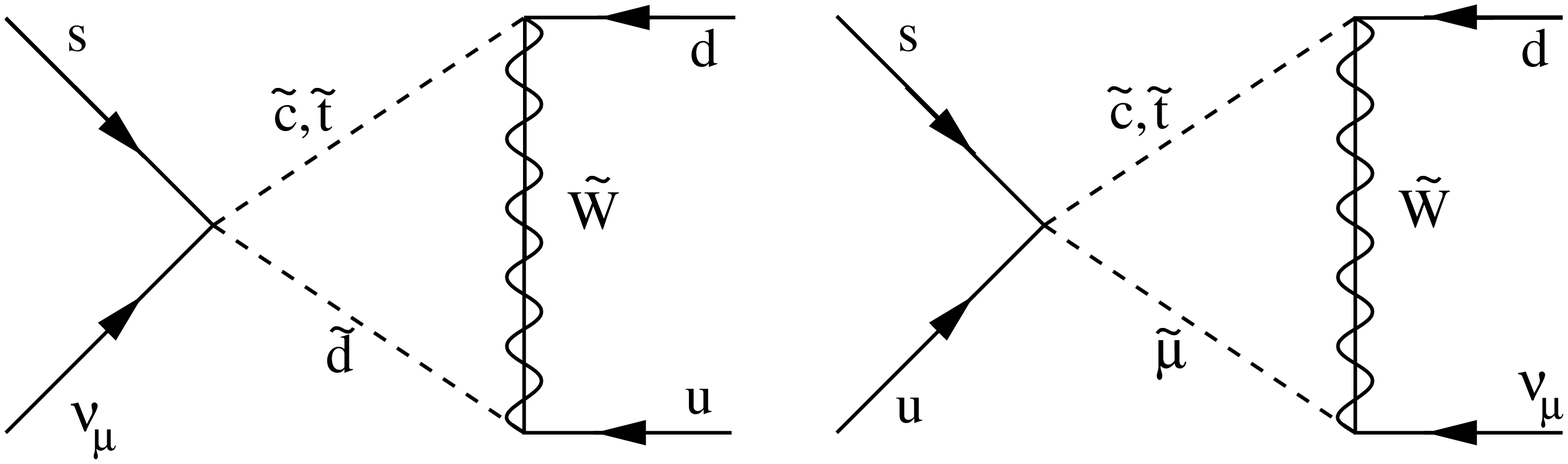}
\caption{Wino-dressing diagrams contributing to $uds\nu$; there are
also diagrams with $s,d$ exchanged.}
\label{fig:udsn}
\end{figure}

 At small $\tanb$ the largest four-fermion operator is $uds\nu$,
induced through dressing of the $CDS\NN_\mu$, $TDB\NN_\mu$ term in the
superpotential by the charged Wino, as in \FFig{udsn}
\cite{rudaz,nath}.  In minimal GUTs the second-generation amplitude is
proportional to $y_c y_s \theta_c^2$ ($y_c$ is the charm quark Yukawa
coupling, {\it etc.}), while the third-generation effect is roughly of
the same order.  Since $y_s y_c\theta_c^2\sim
(\eps_2\zeta)(\eps_2^2)\tc^2 \sim \eps^3\tc^2\zeta$, the coefficient
of \udsn\ saturates its upper bound in the Table.  Similar statements
apply to $udd\nu$, with an additional factor of $\tc$.  Interference
effects between second- and third-generation diagrams [which
unfortunately depend on the otherwise-unmeasurable phases $\sigma_i$
in \Eref{dimfive}] might suppress either $p\rarr K^+\bar\nu$ or
$p\rarr \pi^+\bar\nu$ \cite{nath} but not both \cite{hmy}. Accounting
for this and for the different hadronic matrix elements and phase
space factors, one finds the branching ratio $\Gamma(p\rarr K^+
\overline{\nu})/\Gamma(p\rarr \pi^+ \overline{\nu})$ tends to lie
between .1 and 1. (Throughout this letter we assume that sfermions of
the same charge are degenerate at the messenger scale of supersymmetry
breaking.) In more general GUTs satisfying \Eref{ansatz}, even this
predictivity may be lost, though the tendency is still the same.  The
rates for these processes grow as $\tanbt2$.

\begin{figure}
\centering
\epsfxsize=3.2in
\hspace*{0in}\vspace*{.2in}
\epsffile{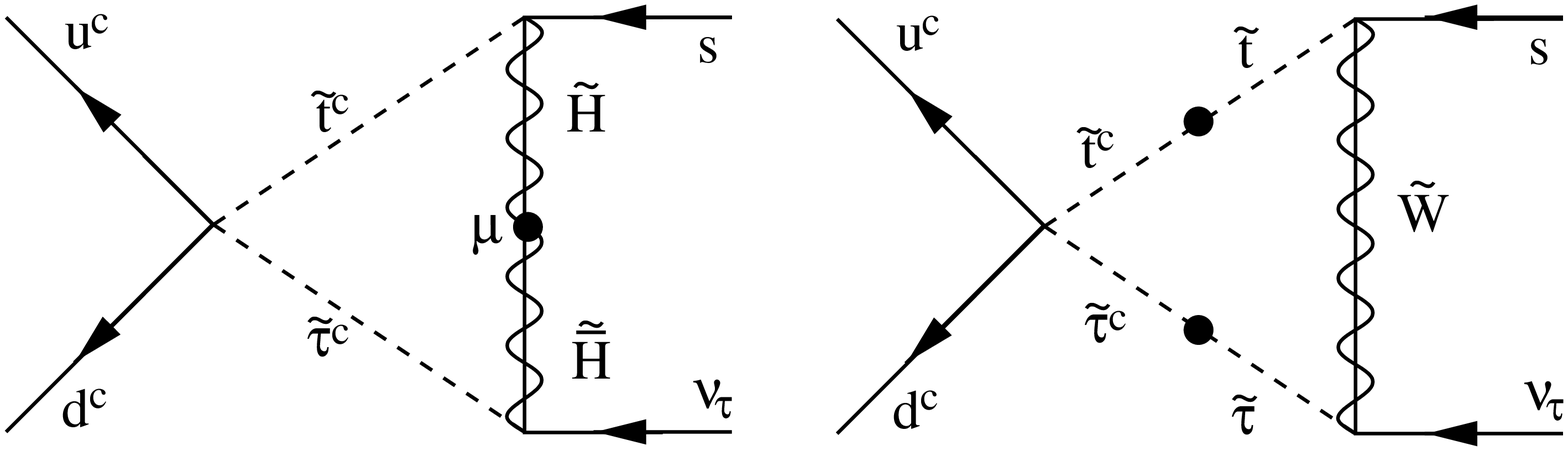}
\caption{The Higgsino- and Wino-dressing diagrams leading
to $u^\cd d^\cd s\nu_{\tau}.$}
\label{fig:ucdcsn}
\end{figure}

At large $\tanb$, the largest operator in these GUTs is
$(u^cd^c)^\dagger(s\nu)$, first discussed in \cite{nathB,lucrab}.
This operator is obtained through Higgsino exchange or through
left-right sfermion mixing, which always entails a factor of the
associated right-handed fermion mass.  The largest effect therefore
comes from changing a third-generation {\it right-handed} squark or
slepton to a light {\it left-handed} quark or lepton. Specifically,
begin with the operator $U^cD^cT^c\tau^c$ in the superpotential, and
then convert the sfermions $\tilde t^c\tilde \tau^c$ to the fermions
$s\nu_{\tau}$.  The Higgsino and Wino dressing diagrams are given in
\FFig{ucdcsn}; there are also diagrams with Higgsino-Wino mixing whose
size is intermediate between them.  The diagrams are proportional to
the coefficient $\hat{\bar{\lambda}}_{1331}$. For the Higgsino
dressing one has the couplings $y_tV_{ts}^*\sim y_t\eps_2$ for the
$\tilde t^c - s$ vertex, $y_\tau$ for the $\tilde\tau^c-\nu_\tau$
vertex.  A similar structure emerges for the Wino diagram but
suppressed by gauge couplings and mixing factors; the Higgsino diagram
usually dominates.  The coefficient of this operator in minimal
$SU(5)$ is proportional to $y_d y_t^2 y_\tau V_{ts}
\sim\eps^3\zeta^2$.  This lies $\eps/\tc$ below its upper bound,
because, as is also the case in many other GUTs,
$\hat{\bar{\lambda}}_{1331}=y_dy_t$ is a factor of $\eps/\tc$ below
the ansatz in \Eref{ansatz}.  This will not be true in non-minimal
GUTs where the matrix $R^{\bar d}$ is non-hierarchical and $R^{\bar
u}_{12}\sim \tc$, as in certain ten-centered models \cite{tenc} where
the bound on this coefficient can be saturated.

The same set of graphs, with only the external down-type quark flavors
changed, gives $(u^c d^c)^\dagger(s \nu_\tau)$, $(u^c s^c)^\dagger(d
\nu_\tau)$ and $(u^c d^c)^\dagger(d \nu_\tau)$.  In minimal $SU(5)$
the three amplitudes are in the ratio $y_dV_{ts},y_sV_{us}V_{td},y_d
V_{td}$; note these ratios are independent of the phases $\sigma_i$,
unlike the $ud_id_j\nu$ case.  The coefficients of $(u^c
d^c)^\dagger(s \nu_\tau)$ and $(u^c s^c)^\dagger(d \nu_\tau)$ are of
opposite sign, and may be of the same order.  (This is slightly
inconsistent with the Table; however, the estimates therein are rough,
and, as noted above, the $(u^c d^c)^\dagger(s \nu_\tau)$ amplitude
does not saturate its bound.)  However, the hadronic matrix elements
of these operators have opposite signs (to see this use
\cite{chw,chadha}) so interference in $p \rightarrow K^+
\overline{\nu}_\tau$ is constructive; also, since $(u^c d^c)^\dagger(s
\nu_\tau)$ has a significantly larger matrix element, its contribution
dominates.  The operator $(u^c d^c)^\dagger(d \nu_\tau)$ leads to $p
\rightarrow \pi^+ \overline{\nu}_\tau$; since its coefficient is in a
known relationship to the other two, a precise prediction for the
branching ratio $\Gamma(p \rightarrow \pi^+ \overline{\nu}_\tau)/
\Gamma(p \rightarrow K^+ \overline{\nu}_\tau)$ is possible,
independent of the supersymmetric spectrum. This predictivity is
retained in models where
$\hat{\bar{\lambda}}_{1331}/\hat{\bar{\lambda}}_{1332}$ is determined.
Even this information is unnecessary in GUTs where
$\hat{\bar{\lambda}}_{1331}$ is as large as allowed in \Eref{ansatz},
since in this case the $(u^c s^c)^\dagger(d \nu_\tau)$ operator is
negligible and only $\hat{\bar{\lambda}}_{1331}$ appears in the
amplitudes.

Since these amplitudes go as $\zeta^2$, the rates for $p\rarr K^+\bar
\nu_\tau,\pi^+\bar\nu_\tau$ grow like $\tanbt4$, leading to short
lifetimes and corresponding strong constraints at large $\tanb$.  We
find that the amplitudes for $uds\nu$ and $(u^cd^c)^\dagger(s\nu)$
become comparable in most GUTs for $\tanb$ somewhere between 3 and 15;
in a minimal supergravity $SU(5)$ GUT (accounting for the different
short-distance renormalizations and hadronic matrix elements of the
two operators) we find this number is of order $9 (m_{\tilde W}/\mu)$,
with large uncertainties from the ratio $y_d/y_s$, the sfermion
spectrum, poorly measured CKM angles, and the third-generation
contribution to the $uds\nu$ operator.\footnote{While this letter was
in preparation, a preprint appeared which studies the
$(u^cd_i^c)^\dagger(d_j\nu)$ operators in the minimal $SU(5)$ GUT
\cite{gotonihei}.}  For $\tanb\sim60$, the rate from
$(u^cd^c)^\dagger(s\nu_\tau)$ dominates that from $uds\nu$ by
$10-1000$.  In those non-minimal GUTs where the bound on
$(u^cd^c)^\dagger(s\nu_\tau)$ is saturated, the amplitude is enhanced
by another factor of $\tc$ and can dominate for even smaller values of
$\tanb$.  The constraints on large $\tanb$ models from these effects,
although discussed in \cite{lucrab}, do not appear to have been fully
incorporated in the literature.

\begin{figure}
\centering
\epsfxsize=1.5in
\hspace*{0in}\vspace*{.2in}
\epsffile{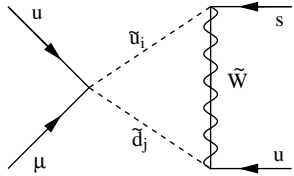}
\caption{The important 
diagrams contributing to $uus\mu$.}
\label{fig:uusl}
\end{figure}

The bounds on $uus\ell$ and $uud\ell$ are comparable to those on
$uds\nu$ and $udd\nu$.  However, in minimal (and some non-minimal)
GUTs, these bounds are not saturated.  From the Table, we see that at
small $\tanb$ we need only consider Wino exchange.  The argument that
$uus\ell$ is highly suppressed \cite{raby,rudaz} is that all
contributions are proportional to $f_1=y_u\sim\eps^4$, which makes the
resulting amplitudes of order $f_1h_2\sim\eps^5$, smaller than the
$\eps^3\tc^2$ bound.  To see this, consider the contribution of $U_i
D_j D_k \NN_r$ in \Eref{dimfive}; the sneutrino must couple to the
Wino, implying $i=1$ and giving a factor of $f_1$.  If instead we use
$U_i D_j U_k E_r$, either $i=1$ or $k=1$.  The former case gives $f_1$
directly, and the latter gives $f_1$ through a unitarity cancellation:
the diagram in \FFig{uusl} is proportional to
\be{CKMcancel}
h_2 V^*_{12} \sum_{i,j=1}^3  V^*_{1j} 
 V_{ij} f_i V^*_{ip} = h_2  f_1 V^*_{12} V_{1p} \ .
\ee
However, the unitarity cancellation in \Eref{CKMcancel} partly fails
due to subtle renormalization group effects.  The operator $U_i D_j
U_k E_r$ is proportional to $V_{ij}$ at the GUT scale, but after
renormalization to low-energy it is no longer proportional to $V_{ij}$
at the weak scale.  This effect is of order $(1-y_t^2/y_f^2)^{1/24}$
\cite{theisenCKM,babuCKM} (here $y_f\sim 1.1$ is the fixed-point value
of $y_t$) or about $0.1$ for $\tanb\sim1.4-3$.  Specifically, consider
the $U_iD_jH_C$ coupling $\hat{F}_{ij}$ (we use the notation of
\cite{babubarr};) naively $\hat{F}_{ij}=f_iV_{ij}$, but in fact
$\hat{F}_{31},\hat{F}_{32}$ differ from this at the weak scale by
$\sim 10 \%$.  The sum over $j$ in \eref{CKMcancel} becomes $h_2
V_{12}^*\sum_j V_{1j}^* \hat{F}_{ij} V_{ip}\sim (0.1) h_2 V_{12}^*
V_{13}^* V_{33} V_{3p}f_3$, which is about a tenth of $h_2f_1$. The
loop factor for this effect can be enhanced if the third-generation
squarks are lighter than those of the first two generations; still,
even if the CKM angles, $f_1$ and the spectrum are at the edge of
their ranges, it can never be as large as the leading contribution.  A
second effect of the same order (if the messenger scale of
supersymmetry breaking scale is high) comes from the mixing and mass
splittings between the down-type squarks.  Both of these effects,
while interesting, are most likely lost in the uncertainties
surrounding $y_u$.

 In many non-minimal GUTs, the unitarity cancellation in
\eref{CKMcancel} simply does not occur.  A sufficient condition for
this is that $\hat{F}_{ij}$ be hierarchical but not precisely equal to
$f_iV_{ij}$. If $\hat{F}_{ij}\sim f_i
\min\{\eps_i/\eps_j,\eps_j/\eps_i\}$ as occurs in many realistic
models of flavor (including those in which higher-dimension operators
or non-minimal Higgs bosons contribute to the up-type quark masses),
the sum $V_{1j}^*\hat{F}_{ij}$ does not equal $f_1\delta_{1i}$;
instead it gives $\tc f_2,\eps\tc f_3$ for $i=2,3$. In this case the
diagrams in \FFig{uusl} with $i=2,3$ are proportional to $h_2 V^*_{12}
(\tc f_2) V^*_{22},h_2 V^*_{12} (\eps\tc f_3) V^*_{32}\sim
\eps^3\tc^2$.  This saturates the bound in the Table, and thus in
these theories the neutrino to charged-lepton branching ratio is order
one at small $\tanb$ (although the hadronic matrix elements favor
neutrinos.)  Enhancements of $F_{ij}$ by factors of
$\tc^{\delta_{i1}+\delta_{j1}}$ do not change this conclusion.

The charged lepton rates from these operators increase only as
$\tanbt2$.  As suggested in the Table, at large $\tanb$ a bigger
contribution to these operators may come from up-type squark mixing in
gluino dressing \cite{goto,bbtwo}; however the branching fraction to
charged leptons remains small due to the large $(u^cd^c)^\dagger
s\nu_\tau$ operator.

 It is clear from the Table that all observed muons should be
left-handed, in contrast to proton decays mediated by dimension-six
operators \cite{weinbergbl,wilczek}.  Although the Table suggests
branching fractions to electrons can be of the same order as those to
muons, in minimal and many non-minimal GUTs they are much smaller
\cite{raby,rudaz}.  In such GUTs the coefficients
$\hat\lambda^e_{ijkr}$ are not roughly independent of $r$, in contrast
to the ansatz \eref{ansatz}. However, the suppression factor is model
dependent, and there are theories (such as ten-centered models
\cite{babubarr,msfermion,tenc}) in which electron and muon decays do
have comparable rates.  The electron-to-muon branching ratios thus are
good probes of flavor physics \cite{kaplan}.

Finally, as evident in the Table, dressings involving gluinos and
neutralinos can be important at small $\tanb$ in their contribution to
the $uds\nu,udd\nu$ operators.  In minimal GUTs these contributions
are suppressed by $y_u$, but in non-minimal GUTs (as in \cite{lucrab})
they may become important.  Gluino dressing is naively subleading due
to the symmetry structure of the dimension-five operators
\cite{raby,vysotskii,pal}, and can only play a role when there is
significant flavor violation \cite{goto,bbtwo}.  In the supersymmetric
basis, flavor violation appears as intergenerational squark mixing.
At small values of tan$\beta$, $\tilde{d_i}-\tilde{d_j}$ mixings are
induced proportional to $y_t^2 V_{3i}V_{3j}^*\sim\eps_i\eps_j$ times
the factor $\gamma$ in the Table.  In minimal $SU(5)$, because neutral
gauginos do not change flavor and intergenerational mixing is small in
the up-squark sector if $\tanb\ll 60$, the $uds\nu$ and $udd\nu$
operators must come from $U_iD_jD_k\NN_r$ with $i=1$; from
\Eref{dimfive} this implies a factor of $f_1=y_u\sim \eps^4$, giving
effects proportional to $y_u y_s \sim \eps^5\ll\eps^3\tc^2$.  But in a
non-minimal GUT, if $\hat{F}_{ij}\neq f_i V_{ij}$, and both $R^u_{12}$
and $R^d_{21}$ are of order $\tc$, then it is possible that
$\hat{F}_{11}\sim {f_2}\tc^2$, giving effects of order $\eps^3\tc^2$,
comparable to the leading Wino dressing contributions \cite{lucrab}.
Dressing by neutralinos ($W_3$-ino and $B$--ino) is similar to the
gluino dressing; however, squark mixing is not required for
neutralinos to contribute to $uds\nu$, so their effects may overshadow
the gluino if $\gamma$ is small \cite{estimates}.  Gluino dressing
also can contribute without mixing to $uds\nu$ if there is substantial
D-term splitting of squark masses \cite{pal}.

\rem{For supersymmetry breaking scenarios in which the first two
generations of sfermions are much heavier than the third
\cite{twoplusone,decoupling,moreminimal} only the
third generation contributes.  The $ud_id_j\nu$ operators
may be slightly reduced, the $(u^cd_i^c)d_j\nu_\tau$ operators are
unchanged, and the $uud_i\ell$ operators saturate their bounds since
the unitarity cancellation in \eref{CKMcancel} fails.  
}

Further details will appear in \cite{estimates}.

\medskip

We thank S. Barr for a preliminary reading of the manuscript.
M.J.S. thanks the Aspen Center for Physics where part of this work was
done.  K.S.B.~was supported in part by DOE grant No. DE-FG02-
90ER-40542 and by funds from the Oklahoma State University.
M.J.S.~was supported in part by National Science Foundation grant NSF
PHY-9513835 and by the W.M.~Keck Foundation.

   \bibliography{gut}        
\bibliographystyle{h-physrev}
\end{document}